\documentclass{ws-ijmpa}
\usepackage[super,compress]{cite}

\usepackage{upgreek}			% Lettere greche maiuscole

\newcommand*{\vb}[1]{\boldsymbol{#1}} %vector bold
\DeclareMathOperator{\tr}{tr}

\usepackage{comment}

\usepackage[dvipsnames, x11names]{xcolor} 
\definecolor{Rosso}{cmyk}{0.3,1,1,0.2}	
\definecolor{Blu}{cmyk}{1,0.6,0,0.2}	
\definecolor{Verde}{cmyk}{1,0.21,1,0.2}

\usepackage{tikz}	
	\tikzset{>=latex}
\usetikzlibrary{calc, intersections, fadings, shadings,%
	positioning, decorations.pathreplacing, arrows, plotmarks, shapes.geometric, arrows.meta, decorations.markings, bending}

\newcommand*{\R}{\mathbb{R}}
\newcommand*{\C}{\mathbb{C}}
\newcommand*{\Z}{\mathbb{Z}}
\newcommand*{\iu}{\mathrm{i}\mkern1mu} %imaginary unit
\newcommand*{\e}{\mathrm{e}} %Euler's number
\newcommand*{\UU}{\mathrm{U}}	
\newcommand*{\SU}{\mathrm{SU}}
\newcommand*{\idm}{\mathrm{I}}

\DeclareMathOperator{\id}{id}

\usepackage{hyperref}
\hypersetup{%
	colorlinks,
	,linkcolor=Rosso
	,citecolor=Blu
	,urlcolor=Verde
}

\begin{document}

\markboth{G. Angelone, P. Facchi and G. Marmo}{Hearing the shape of a quantum boundary condition}

%\catchline{}{}{}{}{}
%
%%%%%%%%%%%%%%%%%%%%%%%%%%%%%%%%%%%%%%%%%%%%%%%%%%%%%%%%%%%%%%%%%%%%

\title{Hearing the shape of a quantum boundary condition}

\author{Giuliano Angelone}

\address{Dipartimento di Fisica, Universit\`a di Bari, I-70126 Bari, Italy\\
INFN, Sezione di Bari, I-70126 Bari, Italy\\
giuliano.angelone@ba.infn.it}

\author{Paolo Facchi}

\address{Dipartimento di Fisica, Universit\`a di Bari, I-70126 Bari, Italy\\
INFN, Sezione di Bari, I-70126 Bari, Italy
}

\author{Giuseppe Marmo}
\address{Dipartimento di Fisica, Universit\`a di Napoli, I-80125 Napoli, Italy\\
INFN, Sezione di Napoli, I-80125 Napoli, Italy}

\maketitle

\begin{history}
\received{Day Month Year}
\revised{Day Month Year}
\end{history}

\begin{abstract}
We study the isospectrality problem for a free quantum particle confined in a ring with a junction, analyzing all the self-adjoint realizations of the corresponding Hamiltonian in terms of a boundary condition at the junction. In particular, by characterizing the energy spectrum  in terms of a spectral function, we classify the self-adjoint realizations in two classes, identifying all the families of isospectral Hamiltonians. These two classes turn out to be discerned by the action of parity (i.e.\ space reflection), which plays a central role in our discussion.
\end{abstract}


\ccode{PACS numbers: 03.65.-w, 03.65.Db}

\section{Introduction}
In  a famous paper of 1966, enticingly entitled ``Can one hear the shape of a drum?'', Mark Kac described an interesting inverse spectral problem involving bounded regions of the plane~\cite{Kac}. 
As it is well known, in any bounded region $\Omega\subset \R^2$ the Laplacian with Dirichlet boundary conditions has a discrete spectrum, say $\{\lambda_n\}_{n\in \mathbb{N}}$, which can in principle always be determined by solving the eigenvalue equation
\begin{equation}
-\Updelta u_n(x,y)=\lambda_n u_n(x,y)\,,\qquad u_n(x,y)|_{\partial\Omega}=0\,,
\end{equation}
assuming that $\Omega$ and its boundary $\partial\Omega$ are sufficiently regular. 
This setting emerges as an idealized model of a drum, described as a two-dimensional vibrating membrane $\Omega$ whose boundary $\partial\Omega$ is kept fixed by a frame, so that the eigenvalues $\lambda_n$ represent its normal frequencies. 
The inverse spectral problem introduced by Kac is thus the following: can we \emph{uniquely} determine the region $\Omega$ just by knowing the corresponding  Laplace eigenvalues $\{\lambda_n\}_{n\in\mathbb{N}}$? 
Or, conversely, can we construct two \emph{isospectral} and non-isometric regions $\Omega_1$ and $\Omega_2$, i.e.\ having different shapes but the same spectrum?

\begin{figure}[tb]
\centering
\begin{tikzpicture}[scale=1.6]
\draw[thick,fill=CadetBlue2] (3,0)--(3,2)--(3.5,2.5)--(4,2)--(4,2)--(3.5,1.5)--(4,1)--(4,0)--(3.5,0.5)--cycle;
\draw[dotted] (3.5,0.5)--(3,1) --(4,1) -- (3.5,0.5) 
(3,1) -- (3.5,1.5)-- (3,2)--(4,2);

\draw[thick,fill=CadetBlue2] (0,0)--(0.5,0.5)--(0,1)--(0,2)--(1,2)--(0.5,1.5)--(1.5,0.5)--(1,0)--cycle;
\draw[dotted] (0.5,0.5) -- (1,0) -- (1,1)--(0.5,0.5) (1,1)--(0,1)--(0.5,1.5)--(0,2);

\node at (-0.5,1) {$\Omega_1$};
\node at (4.5,1) {$\Omega_2$};
\end{tikzpicture}
\caption{A pair of isospectral polygons, constructed with seven half-square tiles differently arranged (the dotted lines are a guide for the eye), see Ref.~\protect\citen{GoWeWo92}  for details.}
\label{fig:isosp}
\end{figure}
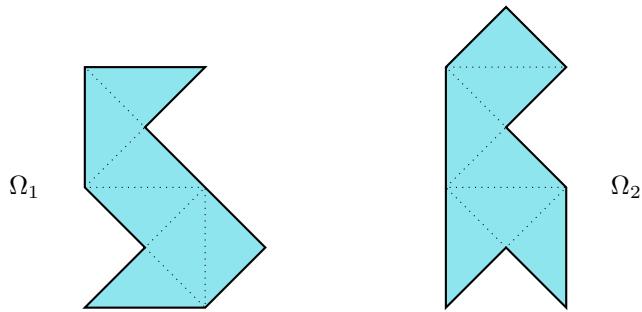

Remarkably, even after more than fifty years, this problem has been only partially solved~\cite{GiTh10, Zel04}. 
A first negative answer has been given in 1992 by C.~Gordon \emph{et al.}~\cite{GoWeWo92}, who constructed an isospectral pair of two-dimensional polygons, depicted in Fig.~\ref{fig:isosp}. 
In the case of two-dimensional domains with a smooth boundary, a positive result can be recovered by requiring some additional symmetries, see e.g. the work of S.~Zelditch~\cite{Zel00}, but the general problem is still unsolved. 
Besides, many related questions (which may generally fall under the umbrella of ``isospectrality'') are still open and object of research. 
Here we briefly mention some interesting topics, involving the isospectrality problem for regions with a fractal boundary~\cite{BrCa86},  relativistic billiards~\cite{DiHu20}, graphs~\cite{GuSm01, Rue15}, photonic systems~\cite{photo21}, and also the experimental construction of some isospectral regions~\cite{SrKu94, EvPi99}.

In this Article we will study an analogous problem, determining which boundary conditions, associated with a free quantum particle in a ring with a junction, have the same energy eigenvalues. 
In other words we fix the shape of the ``drum'' (which will be a one-dimensional ring) and we ask whether it is possible to distinguish different quantum boundary conditions just by the energy spectrum of the particle.

Apart from being the simplest nontrivial mathematical model to perform our analysis, the one-dimensional problem can be implemented in a superconducting quantum interference device (SQUID), a superconducting ring where different boundary conditions can be obtained by a flux-loop tunable junction~\cite{Vion,Poletto,Paauw}. 

From a physical perspective the inverse spectral problem might have applications in quantum metrology with SQUIDs~\cite{Friedman}, atoms in cavities~\cite{Haroche}, and ions and atoms in magnetic traps~\cite{Paul}.
  
We remark that, although the physical configuration is fixed,  the topology of the quantum system is not, being determined by the  boundary conditions: while non-local boundary conditions effectively describe a particle in a ring, local conditions model instead a particle confined in a segment (i.e. a one-dimensional box), see e.g.\ Refs.~\citen{topchange, AIM05,SWX12,IBP15,trotter,newboundaries} for further details.

The paper is organized as follows. In Sec.~\ref{sec:qbc} we introduce the system and the related Hamiltonian, describing all the allowed boundary conditions which ensure the self-adjointness of the Hamiltonian.
Then in Sec.~\ref{sec:specfun} we characterize the spectrum of the Hamiltonian in terms of the roots of a certain function. 
In Sec.~\ref{sec:symm}, finally, after discussing the dependence of the spectrum on the boundary conditions, and interpreting the result by analyzing the behavior of the system under a parity transformation,  we determine the geometric structure of the space of quantum boundary conditions and its relation with the space of Hamiltonians.

\section{Quantum boundary conditions on a ring}\label{sec:qbc}
We are interested in a free quantum particle in a ring with a junction. 
This system, which is essentially one-dimensional, is described by the kinetic-energy operator
\begin{equation}\label{eq:HU}
H=-\frac{\hbar^2}{2m}\frac{\mathrm{d}^2}{\mathrm{d}x^2}
\end{equation}
acting on the Hilbert space $L^2(-\ell/2,\ell/2)$, where $\ell$ denotes the length of the ring and $m$ is the mass of the particle. Equation~(\ref{eq:HU}) describes the action of $H$ in the ring away from the junction. In order to generate a well-defined dynamics, the Hamiltonian $H$ should be equipped with suitable boundary conditions, which specify the behavior of the particle at the junction. In quantum mechanics the possible behavior at the boundary, encoded in the domain $\mathcal{D}(H)$ of $H$, cannot be arbitrary, but is constrained by a basic principle: $H$ must be \emph{self-adjoint}, i.e.\ $\mathcal{D}(H)=\mathcal{D}(H^\dagger)$ and $H=H^\dagger$. Indeed, self-adjointness is a necessary and sufficient condition for the operator to have a purely real spectrum and to generate a unitary dynamics. 

Different domains correspond to different behavior of the particle at the junction and thus give rise to different dynamics. All the self-adjoint realizations  of the kinetic energy $H$ are known to be in one-to-one correspondence with the set of $2\times 2$ unitary matrices $U\in\UU(2)$~\cite{AIM05, AIM15}. Each of these realizations, which we henceforth denote by $H_U$, is defined on the domain
\begin{equation}
\mathcal{D}(H_U)=\{\psi\in \mathcal{H}^2(-\ell/2,\ell/2):(\idm-U)\Psi=\iu(\idm+U)\Psi'\}\,,
\end{equation}
where $\mathcal{H}^2(-\ell/2,\ell/2)$ denotes the space of wave functions with square-integrable first and second derivative (second Sobolev space) on $[-\ell/2,\ell/2]$, $\idm$ is the $2\times 2$ identity matrix, and 
\begin{align}
\Psi\equiv
\left(\begin{array}{@{}c@{}}
\psi(-\ell/2)\\ \psi(\ell/2)
\end{array}\right)\,, &&
%\qquad\textnormal{and}\qquad 
\Psi' \equiv \ell_0
\left(\begin{array}{@{}c@{}}
-\psi'(-\ell/2)\\ \psi'(\ell/2)
\end{array}\right)\,,
\end{align}
are the boundary values, with $\ell_0$ an arbitrary scale length which we take as $\ell_0=\ell$ for convenience. In other words, to each unitary matrix $U$ there corresponds the \emph{quantum boundary condition}
\begin{equation}\label{eq:be}
(\idm-U)\Psi=\iu(\idm+U)\Psi'\,,\qquad U\in\UU(2)\,,
\end{equation}
which in turn prescribes a linear relation between the vectors of boundary data $\Psi$ and $\Psi'$. 

In the following we are interested in the relationship between the spectrum $\sigma(H_U)$ of the Hamiltonian $H_U$ and its corresponding quantum boundary condition $U$. Before moving on, let us spend a few words on the parametrization of the unitary group $\UU(2)$, as it will be crucial to determine the geometry of the boundary conditions.

\subsection{Parametrization of the unitary group}
The unitary group $\UU(2)$ splits over $\SU(2)$ as $\UU(2)=\SU(2)\rtimes\UU(1)$, by the  map 
\begin{align}
	\SU(2)\times\UU(1)\to\UU(2)\,, &&
	(M, \e^{\iu \eta})\mapsto M\left(\begin{array}{@{}cc@{}}
	\e^{\iu\eta} & 0 \\ 0 & 1
	\end{array}\right)\,,
\end{align}
which usefully captures the group structure of $\UU(2)$. In the following, however, we are more interested in the manifold structure of this Lie group. For our purposes it will be thus more convenient to use a parametrization via the  map 
\begin{align}  \label{eq:covering}
\SU(2)\times\UU(1)\to\UU(2)\,, &&  (M,\e^{\iu \eta})\mapsto \e^{\iu \eta} M\,.
\end{align}
This map represents a double covering of $\UU(2)$, since both  $(M,\e^{\iu \eta})$ and $(-M,-\e^{\iu \eta})$ are mapped to the same element $\e^{\iu\eta}M$,
 and this means that $\UU(2)$ is given topologically by the quotient $(\SU(2)\times \UU(1))/\mathbb{Z}_2$, see \ref{sec:app} for further details. 
 In any case, the relevant point here is that by restricting the angle $\eta$ to the interval $[0,\pi[$, we get a one-to-one parametrization of $\UU(2)$.

For what concerns the parametrization of the $\SU(2)$ part, a generic element $M\in \SU(2)$ can be represented by the matrix~\cite{Wig59}
\begin{equation}
M= \left(\begin{array}{@{}cc@{}}
	a & b^{\ast} \\ -b & a^{\ast}
\end{array}\right)\,,\qquad \left|a\right|^2+\left|b\right|^2=1\,,
\end{equation}
the constraint on the complex numbers $a$ and $b$ implying that $\det(M)=1$. Equivalently, we can express $M$ in terms of four real parameters $m_0$ and $\vb{m}=(m_1,m_2,m_3)$ subjected to the $S^3$ constraint $m_0^2+|\vb{m}|^2=1$, that is:
\begin{equation}\label{eq:Upar}
M=m_0 \idm+\iu\vb{m}\cdot \vb{\sigma}=
\left(\begin{array}{@{}cc@{}}
	m_0+\iu m_3 & m_2+\iu m_1  \\ -m_2+\iu m_1 & m_0-\iu m_3
\end{array}\right)\,,
\end{equation}
where $\vb{\sigma}=(\sigma_x, \sigma_y, \sigma_z)$ is the vector of the Pauli matrices. In the following we will need the following quantities in terms of  the  covering map~\eqref{eq:covering}:
\begin{equation}\label{eq:Uquant}
\det (U)=\e^{\iu 2 \eta}\,,\qquad \tr (U)= 2\,\e^{\iu\eta}m_0\,,\qquad\text{and}\qquad \tr (U\sigma_x)=2\iu\e^{\iu\eta}m_1\,.
\end{equation}

\section{Spectrum via a spectral function}\label{sec:specfun}
In this section we characterize the energy spectrum $\sigma(H_U)$ in terms of a suitable function. Let us introduce the notation
\begin{equation}
\Psi_{\pm}\equiv \Psi\pm\iu\Psi'=
\left(\begin{array}{@{}c@{}}
	\psi(-\ell /2)\mp\iu \ell\psi'(-\ell /2)\\ \psi(\ell /2)\pm\iu \ell\psi'(\ell /2)  
\end{array}\right)\,,
\end{equation}
so that Eq.~\eqref{eq:be} can be compactly written as $\Psi_-=U\Psi_+$. The eigenvalue equation $(H_U-E)\psi(x)=0$, namely
\begin{align}\label{eq:HUE}
	\psi''(x)+\epsilon\psi(x)=0\,,&& \epsilon\equiv 2mE/\hbar^2\in \R\,,
\end{align}
has a general solution of the form
\begin{equation}\label{eq:gensol}
%\psi(x; k)=c_1\psi_1(x; k)+c_2\psi_2(x; k)\,,
\psi(x; \epsilon)=c_1\psi_1(x; \epsilon)+c_2\psi_2(x; \epsilon)\,,
\end{equation}
where $c_1, c_2\in\C$. %and where we introduced the \emph{wave-number} $k=\sqrt{\epsilon}$. 
Observe that %this latter quantity can be either real or purely imaginary, respectively when $\epsilon$ is 
$\epsilon$ can be either positive or %strictly 
negative. Inserting this solution in the expression for $\Psi_\pm$ we get
\begin{equation}
%\Psi_\pm=A_\pm(k)\left(\begin{array}{@{}c@{}}
%c_1 \\ c_2
%\end{array}\right)
\Psi_\pm=A_\pm(\epsilon)\left(\begin{array}{@{}c@{}}
c_1 \\ c_2
\end{array}\right) \, ,
\end{equation}
where
\begin{equation}
A_\pm(\epsilon)= \left(\begin{array}{lll}
\psi_1\bigl(-\frac{\ell}{2};\epsilon\bigr)\mp\iu \ell\psi_1'\bigl(-\frac{\ell}{2};\epsilon\bigr) & \quad & \psi_2\bigl(-\frac{\ell}{2};\epsilon\bigr)\mp\iu \ell\psi_2'\bigl(-\frac{\ell}{2};\epsilon\bigr) \\ \\
\psi_1\bigl(+\frac{\ell}{2};\epsilon\bigr)\pm\iu \ell\psi_1'\bigl(+\frac{\ell}{2};\epsilon\bigr) & \quad & \psi_2\bigl(+\frac{\ell}{2};\epsilon\bigr)\pm\iu \ell\psi_2'\bigl(+\frac{\ell}{2};\epsilon\bigr)
\end{array}\right)\, .
\end{equation}
Thus we obtain that, for nontrivial $\Psi_\pm$,
\begin{equation}\label{eq:PUP}
\Psi_-=U\Psi_+\qquad\Leftrightarrow\qquad F_U(\epsilon)\equiv \det[B(\epsilon)-U]=0\, ,
\end{equation}
where 
\begin{equation}
  B(\epsilon)\equiv A_-(\epsilon)A_+^{-1}(\epsilon)\,. 
\end{equation}
The function $F_U(\epsilon)$ is called the \emph{spectral function} of the system, and characterizes the spectrum of $H_U$ in terms of its roots:
\begin{equation}
\sigma(H_U)=\{E\in\R : F_U(2mE/\hbar^2)=0\} = \frac{\hbar^2}{2 m} \ker(F_U)\,.
\end{equation}
We can manipulate the expression of $F_U(\epsilon)$ by making use of the relation 
\begin{equation}
\det(M-N)=\det(M)+\det(N) + \tr(MN) - \tr(M)\tr(N)\,,
\end{equation}
which is generally true for any pair of $2\times 2$ matrices~\cite{BFV01}, obtaining that
\begin{equation}\label{eq:FU}
F_U(\epsilon)=\det(B(\epsilon))+\det(U)+\tr(B(\epsilon)U)-\tr(B(\epsilon))\tr(U)\,.
\end{equation}

Before computing the explicit expression of $F_U(\epsilon)$, some remarks are in order. To obtain the equivalence~\eqref{eq:PUP}, and in particular to define $B(\epsilon)$, the matrix $A_+(\epsilon)$ has to be invertible: in the following we will explicitly show that this is indeed the case.  Besides, one may be tempted to define a different spectral function 
\begin{equation}\label{eq:GU}
\tilde{F}_U(\epsilon)\equiv\det[A_-(\epsilon)-UA_+(\epsilon)]\,.
\end{equation}
Since $\ker(\tilde{F}_U)=\ker(F_U)$, for our purposes the two functions are indeed equivalent.
As it turns out however, $\tilde{F}_U(\epsilon)$ presents some subtleties regarding the point $\epsilon=0$, namely the fact that $\tilde{F}_U(0)\neq \lim_{\epsilon\to 0} \tilde{F}_U(\epsilon)$. Instead, $F_U(\epsilon)$ is continuous for all $\epsilon\in\R$.%(at least) for each $k\in\R\cup\iu\R$.

\subsection{Solution of the eigenvalue equation}\label{sec:ee}
To find the general solution~\eqref{eq:gensol}, and thus to obtain the explicit expression of $B(\epsilon)$, it is convenient to distinguish the solutions with zero energy (i.e., the \emph{zero modes}) from the others.

\subsubsection{Zero modes}
When $\epsilon=0$ % the secular equation associated with Eq.~\eqref{eq:HUE}, namely $\lambda^2=0$, admits only the degenerate pair of roots $\lambda_\pm=0$. The solution reads thus
the eigenfunctions read
\begin{equation}
\psi(x; 0)= c_1 +c_2 x/\ell \,,	
\end{equation}
from which we get 
\begin{align}
A_{\pm}(0)=\left(\begin{array}{@{}cc@{}}
1\, & + 1/2\pm \iu \\ 1\, & -1/2\mp \iu
\end{array}\right)\,,&&
%\qquad\text{and}\qquad 
B(0)=\frac{1}{1+2\iu}\left(\begin{array}{@{}cc@{}}
1 & 2\iu \\ 2\iu & 1
\end{array}\right)\,.
\end{align}

\subsubsection{Solutions with nonzero energy}
When $\epsilon\neq 0$ %we have two distinct roots of the secular equation $\lambda^2+\epsilon=0$, given by $\lambda_\pm=\pm\iu k\neq 0$. The corresponding solution takes then the form
the solution takes the form
\begin{equation}
		\psi(x; \epsilon)= c_1 \e^{\iu k x}+c_2 \e^{-\iu k x}\,, \qquad (\epsilon\neq 0),
\end{equation}
where
\begin{equation}
	k = k(\epsilon) \equiv \begin{cases}
		\sqrt{\epsilon}, & \epsilon\geq 0,\\
		\iu\sqrt{|\epsilon|}, & \epsilon < 0,
	\end{cases}
	\label{eq:k(eps)}
\end{equation}
is the wave number. 
By a straightforward calculation we get that
\begin{equation}
A_\pm(\epsilon)=\left(\begin{array}{@{}ll@{}}
(1\pm k\ell )\e^{\iu k\ell /2} \quad & (1\mp k\ell )\e^{-\iu k\ell /2} \\ (1\mp k\ell )\e^{-\iu k\ell /2} \quad & (1\pm k\ell )\e^{\iu k\ell /2}
\end{array}\right)
\end{equation}
and that
\begin{equation}
B(\epsilon)=\frac{2\iu}{\det [A_+^{-1}(\epsilon)]}\left(\begin{array}{@{}cc@{}}
[1-(k\ell )^2]\sin(k\ell ) \quad & 2\iu k\ell  \\  2\iu k\ell  & [1-(k\ell )^2]\sin(k\ell )
\end{array}\right)
\end{equation}
where $\tfrac{1}{2\iu} \det [A_+^{-1}(\epsilon)]=[1+(k\ell )^2]\sin(k\ell )+2\iu k\ell \cos(k\ell )$. Remarkably, although the matrices
\begin{equation}
\lim_{\epsilon\to 0} A_\pm(\epsilon)=\left(\begin{array}{@{}cc@{}}1 & 1 \\ 1 & 1
\end{array}\right)
\end{equation}
are not invertible and different from $A_{\pm}(0)$ (which are instead invertible), the limit matrix
\begin{equation}
\lim_{\epsilon\to 0} B(\epsilon)=\frac{1}{1+2\iu}\left(\begin{array}{@{}cc@{}}1 & 2\iu \\ 2\iu & 1
\end{array}\right)=A_{-}(0)A_{+}(0)^{-1} = B(0)
\end{equation}
is well-defined. This means that, unlikely $A_{\pm}(\epsilon)$, the matrix valued function $B(\epsilon)$ is continuous at $\epsilon=0$. As we already observed, this is exactly the unwanted feature affecting the alternative spectral function $\tilde{F}_U(\epsilon)$ in Eq.~\eqref{eq:GU}.

\subsection{Spectral function}
By gathering the previous results, we obtain the expression
\begin{equation}
B(\epsilon)=%\left(\begin{array}{@{}cc@{}}a(k) & b(k) \\ b(k) & a(k)\end{array}\right)=
a(\epsilon) \idm + b(\epsilon) \sigma_x \,,
\end{equation}
defined for all $\epsilon\in\R$
%each $k\in\R\cup \iu\R$ 
(eventually by a limit when $\epsilon=0$), where 
\begin{align}
a(\epsilon)&\equiv\frac{[1-(k\ell )^2]\sin(k\ell )}{[1+(k\ell )^2]\sin(k\ell )+2\iu k\ell \cos(k\ell )}\stackrel{\epsilon\to 0}{\longrightarrow}\frac{1}{ 1 + 2\iu }\,,\\
b(\epsilon)&\equiv\frac{2\iu k\ell }{[1+(k\ell )^2]\sin(k\ell )+2\iu k\ell \cos(k\ell )}\stackrel{\epsilon\to 0}{\longrightarrow}\frac{2\iu }{ 1 + 2\iu }\,,
\end{align}
and $k=k(\epsilon)$ is given in~\eqref{eq:k(eps)}.
Since $\tr(B(\epsilon))=2a(\epsilon)$ and $\det(B(\epsilon))=a^2(\epsilon)-b^2(\epsilon)\equiv c(\epsilon)$, with
\begin{equation}
%c(\epsilon)=\frac{[1+(k\ell )^2]\tan(k\ell )-2\iu k\ell }{[1+(k\ell )^2]\tan(k\ell )+2\iu k\ell } \stackrel{\epsilon\to 0}{\longrightarrow} \frac{1-2\iu }{1+2\iu} \,,
c(\epsilon)=\frac{[1+(k\ell )^2]\sin(k\ell )-2\iu k\ell \cos(k\ell ) }{[1+(k\ell )^2]\sin(k\ell )+2\iu k\ell \cos(k\ell ) } \stackrel{\epsilon\to 0}{\longrightarrow} \frac{1-2\iu }{1+2\iu} \,,
\end{equation}
by using Eq.~\eqref{eq:FU} we finally obtain the explicit expression of the spectral function
\begin{align}
F_U(\epsilon)&=\det(U)-a(\epsilon)\tr(U)+b(\epsilon)\tr(U\sigma_x)+c(\epsilon)\label{eq:FU1} \\
&=\e^{2\iu\eta}-2\e^{\iu\eta}m_0a(\epsilon)+2\iu \e^{\iu\eta}m_1 b(\epsilon)+c(\epsilon)\,,\label{eq:FU2}
\end{align}
where in the second line we have used Eq.~\eqref{eq:Uquant}. 

Before proceeding with our analysis, two comments are in order. Notice that the system admits zero modes if and only if $F_U(0)=0$, which in our parametrization~\eqref{eq:Upar} of $U\in\UU(2)$ gives the condition
\begin{equation}\label{eq:FU0}
\cos(\eta)-2\sin(\eta)=m_0+2m_1\,,
\end{equation}
see also Ref.~\citen{FT00} for a related result. In particular, when $m_0^2+m_1^2=1$ so that $m_0=\cos(\theta)$ and $m_1=\sin(\theta)$ for a $\theta\in[0,2\pi[$, the above condition can be rewritten as
\begin{align}
\cos(\eta+\eta_0)=\cos(\theta-\eta_0)\,,&& \eta_0\equiv \arctan(2)\,,
\end{align}
thus admitting the two solutions $\theta_1(\eta)=\eta+2\eta_0$ and $\theta_2(\eta)=-\eta$, whose only intersection occurs at the point $\eta=-\eta_0$.

Moreover, notice that the full information about $\sigma(H_U)$ is embedded in a single continuous function, that is $F_U(\epsilon)$. However, the alternative discontinuous spectral function $\tilde{F}_U(\epsilon)$ given in Eq.~\eqref{eq:GU} has usually a simpler expression for $\epsilon\neq 0$~\cite{FT00,BFV01,AIM15}.

\section{Symmetries and geometry of the spectral space}\label{sec:symm}
The  spectral function $F_U(\epsilon)$ given in Eq.~\eqref{eq:FU2}, and consequently the energy spectrum $\sigma(H_U)$, depends only on \emph{three} of the four independent real parameters parametrizing the unitary $U\in\UU(2)$, namely on $\eta$, $m_0$ and $m_1$, but not on $m_2$ \emph{or} $m_3$ (recall that, since $m_0^2+\vb{m}^2=1$, once $m_0$ and $m_1$ have been fixed only one parameter between $m_2$ and $m_3$ is actually independent).  This result was obtained in Refs.~\citen{FT00,BFV01}. Besides, a similar result was obtained in Ref.~\citen{TFC01}  for a particle in a segment with Dirichlet boundary conditions and a point interaction at the middle point. 

In the remaining part of this paper, we will better understand the symmetries of the energy spectrum by analyzing  the action of the parity operator (and of the $\UU(1)$ group it generates) on the Hamiltonians $H_U$, and we will also clarify the geometric structure of the space of  the quantum boundary conditions.

\subsection{Action of the parity transformation}
The fact that $\sigma(H_U)$ is independent of $m_2$ (or $m_3$) is a consequence of a symmetry of the spectral function: for any 
$\delta\in\mathbb{R}$
we have that 
\begin{equation}
	F_U(\epsilon)=F_{U_\delta}(\epsilon),
\end{equation}
 where
\begin{equation}\label{eq:Udelta}
U_\delta\equiv \e^{\iu\delta\sigma_x}U\e^{-\iu\delta\sigma_x}\,.
\end{equation}
One can easily verify that the above $\UU(2)$ transformation $U\mapsto U_\delta$ is indeed the most general one which simultaneously preserves the values of $\det(U)$, $\tr(U)$ and $\tr(U\sigma_x)$ appearing in the expression~\eqref{eq:FU1} of the spectral function. 

Moreover, observe that this symmetry, which can be regarded as a symmetry of the \emph{boundary}, is ultimately due to a symmetry of the \emph{bulk} associated with the one-parameter unitary group 
\begin{equation}
\mathcal{P}_{\delta}\equiv \e^{\iu\delta P}=\cos(\delta) \idm+\iu\sin(\delta) P\,,\qquad \delta\in\mathbb{R}\,,
\end{equation}
generated by the parity operator 
\begin{equation}
P\colon \psi(x)\mapsto\psi(-x)
\end{equation}
on $L^2(-\ell /2,\ell /2)$. 
For each $\delta\in\mathbb{R}$ the action of $\mathcal{P}_{\delta}$ leaves formally invariant the Hamiltonian~\eqref{eq:HU}, that is the Laplacian, but  changes its domain according to
\begin{equation}\label{eq:PHP}
\mathcal{P}_{\delta} H_U \mathcal{P}_{\delta}^{\dagger}=H_{U_\delta}\,,
\end{equation}
with $U_\delta$ given by Eq.~\eqref{eq:Udelta}. 

We thus conclude that $\mathcal{P}_{\delta}$, although strictly speaking is not a symmetry of $H_U$ (as in general $[\mathcal{P}_{\delta}, H_U]\neq 0$), establishes the unitary equivalence of two \emph{generally distinct} self-adjoint realizations of the same operator. The unitary $\mathcal{P}_{\delta}$ connects two different physical situations, corresponding to two distinct interactions of the particle with the boundary, and thus to two different quantum dynamics.

As a consequence, for each $\delta\in\R$ we get  the following \emph{isospectral relation}:
\begin{equation}\label{eq:isospectrality}
\sigma(H_U)=\sigma(H_{\e^{\iu\delta\sigma_x}U\e^{-\iu\delta\sigma_x}})\,.
\end{equation}
Thus, the corresponding relation associated with the generator $P$ which was already observed in similar settings~\cite{TFC00,CFT01,TFC01}, namely
\begin{equation}
PH_UP^\dagger=H_{\sigma_x U\sigma_x}\,,
\end{equation}
constitutes only the particular case $\delta=\pi/2$ of our Eq.~\eqref{eq:PHP}. On the other hand, from the relation $[\mathcal{P}_\delta,H_U]=\iu\sin(\delta)[P,H_U]$
we obtain that $P$ is a symmetry of $H_U$, in the sense that $[P,H_U]=0$, if and only if $\mathcal{P}_{\delta}$ is a symmetry of $H_U$ for any non-trivial $\delta\neq 0$. 

Moreover, we stress  that $P$ is not a symmetry of $H_U$ for all $U\in\UU(2)$. From a mathematical point of view this is due to the fact that the commutator $[P,H_U]$ actually depends on the domain of $H_U$, that is on the boundary conditions. From a physical point of view we may equivalently argue that imposing  a suitable boundary condition can possibly induce a spontaneous symmetry breaking, see e.g.\ Refs.~\citen{Ca77, AIM15} for a  discussion.

In any case, the relation~\eqref{eq:PHP} can be used to determine the full family of self-adjoint Hamiltonians which are actually symmetric under $\mathcal{P}_\delta$ (or, equivalently, under $P$). By solving Eq.~\eqref{eq:Udelta} for $U=U_\delta$, we find that such operators belong to the $(\UU(1)\times \UU(1))/\mathbb{Z}_2$ subgroup  of $\UU(2)$ described by the family of matrices
\begin{equation}\label{eq:Uparity}
U(\eta,\theta)\equiv \e^{\iu (\eta \idm+\theta \sigma_x)}=\e^{\iu\eta}\left(\begin{array}{@{}cc@{}}
	\cos(\theta) & \iu\sin(\theta)  \\ \iu\sin(\theta) & \cos(\theta)
\end{array}\right)\,,\qquad \eta\in[0,\pi[\,,\quad \theta\in[0,2\pi[\,.
\end{equation}
They form the unitary group of the Abelian C$^*$-algebra generated by $\sigma_x$. In terms of the parametrization~\eqref{eq:covering}--\eqref{eq:Upar}, $U(\eta,\theta)$ is obtained by setting $m_0=\cos(\theta)$, $m_1=\sin(\theta)$ and $m_2=m_3=0$.

\begin{figure}[tp]
\centering
\tikzset{
dot/.style={circle, draw, fill=red, inner sep=0, minimum width=4pt, outer sep=0pt},
partial ellipse/.style args={#1:#2:#3}{insert path={+ (#1:#3) arc (#1:#2:#3)}}
}
\begin{tikzpicture}
\node at (-5.5,1.5) {(a)};
\node at (-5.5,-2.3) {(b)};

\begin{scope}[scale=0.8]
\node at (-4.8,1.75) {$S^1$};
\draw[very thick,  -|] (-3.5,0)  arc (0:360:1.4) node[black, right, xshift=5pt] {$0\sim 2\pi$};
 
\draw[thick, ->] (-1.35,0)--(-0.15,0) node[midway, above] {$\Z_2$};

\draw[very thick, ] (3.2,0)  arc (0:360:1.4);

\node[dot] (A) at ($(1.8,0)+ (0:1.4)$)  {};
\node[dot] (B) at ($(1.8,0)+ (60:1.4)$)  {};
\node[dot] (C) at ($(1.8,0)+ (120:1.4)$)  {};
\node[dot] (D) at ($(1.8,0)+ (180:1.4)$)  {};
\node[dot] (E) at ($(1.8,0)+ (240:1.4)$)  {};
\node[dot] (F) at ($(1.8,0)+ (300:1.4)$)  {};

\draw [<->, gray] (A)--(D); 
\draw [<->, gray] (B)--(E);
\draw [<->, gray] (C)--(F); 

\node at (3.73,0) {$\cong$};

\draw[very thick, -|] (5.6,0)  arc (0:360:0.7)  node[black, right, xshift=5pt] {$0\sim \pi$}; 
\node at (5,1.05) {$S^1\!/\Z_2$};
\end{scope}
% % % % % % % % % % % % % % % % % % % %
\begin{scope}[xscale=0.6, yscale=1, yshift=-110pt, xshift=-60pt]
\fill[fill=CadetBlue3] (0,0) arc (0:360:1);
\draw[very thick, postaction={decorate}, decoration={markings, mark=at position 0.365 with {\arrow{To}}}, decoration={markings, mark=at position 0.415 with {\arrow{To}}}] (-1,1) arc (90:270:1);
\draw[very thick, ->, red] (2,0.7) [partial ellipse=-60:240:6.4cm and 0.8cm];
\fill[CadetBlue1] (-1,-1) arc (-90:90:1) --++ (6,0) arc (90:-90:1) -- cycle;

\draw[very thick, ->, >={latex[bend]}]  (-3,0.15) arc (340:0:0.3) node[pos=0.2, below, black] {$\pi$};

\draw[very thick, dashed, postaction={decorate}, decoration={markings, mark=at position 0.39 with {\arrow{To}}},] (5,1) arc (90:270:1);
\draw[very thick, postaction={decorate}, decoration={markings, mark=at position 0.35 with {\arrow{To}}},decoration={markings, mark=at position 0.4 with {\arrow{To}}}] (5,-1) arc (-90:90:1);

\draw[very thick] (-1,1) --++ (6,0) (-1,-1)--++ (6,0);

\draw[very thick, postaction={decorate}, decoration={markings, mark=at position 0.375 with {\arrow{To}}},] (-1,-1) arc (-90:90:1);

\draw[very thick, ->] (-1.5,-1.3) -- (6.5,-1.3) node[right] {$\eta$};
\draw (-1,-1.2) -- (-1,-1.4) node[below] {$0$};
\draw (5,-1.2) -- (5,-1.4) node[below] {$\pi$};
\end{scope}

\end{tikzpicture}

\caption{(a) The circle $S^1=\{\e^{\iu\eta}: \eta\in[0,2\pi[\}$ can be constructed by imposing the equivalence relation $\eta\sim \eta+2\pi$ on the real line $\R$, namely $S^1=\R/(2\pi \Z)$. Further imposing $\eta\sim \eta+\pi$ corresponds to identify its antipodal points by acting with $\Z_2$, obtaining thus another circle with half the radius. Note that on $S^1$ another $\Z_2$-action can be given, by identifying $\eta\sim -\eta$, and this results in the \emph{orbifold} segment $[0,\pi]$, see e.g.\ Ref.~\protect\citen{orbi1}. (b) The spectral space $\Sigma\cong D\times (S^1/\mathbb{Z}_2)$ can be constructed by gluing the two bases of the solid cylinder $D\times [0,\pi]$ after applying a global twist of angle $\pi$,  identifying thus the edge segments with the same number (one or two) of arrows.}
\label{fig:quotient}
\end{figure}
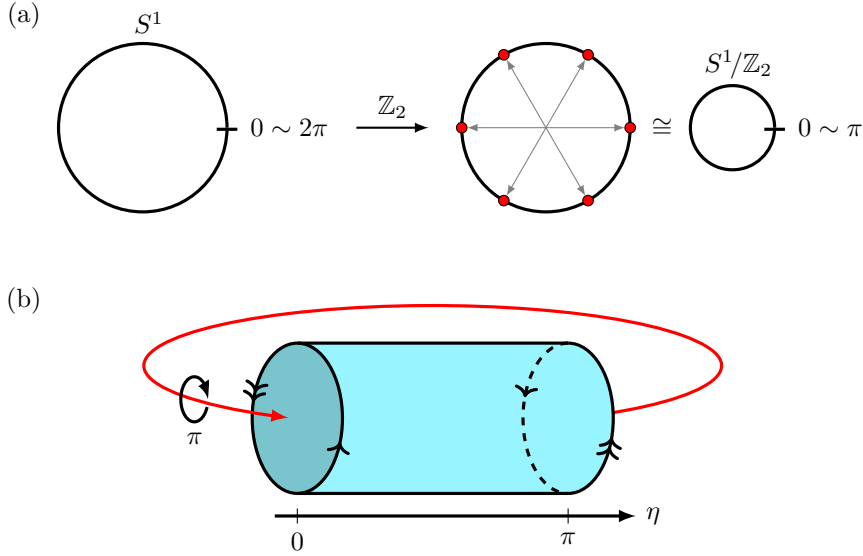

\subsection{A twist in the spectral space}\label{sec:twist}
Let us consider the space of \emph{distinct} spectra, that is the set
\begin{equation}
\Sigma\equiv\{\sigma(H_U) \,:\, U\in\UU(2)\}\,,
\end{equation}
where, according to Eqs.~\eqref{eq:Udelta} and~\eqref{eq:PHP}, the unitary matrices $U$ and $U_\delta$  correspond to the same element $\sigma(H_U)=\sigma(H_{U_\delta})$ of $\Sigma$ for any $\delta\in\R$. 
From our previous discussion it follows that the \emph{spectral space} $\Sigma$ can be parametrized by $\eta$, $m_0$ and $m_1$.

Therefore, seen as a manifold, $\Sigma$ has the same structure of $D\times (S^1\!/\mathbb{Z}_2)$, which is a solid torus  \emph{twisted} by an angle $\pi$. Here, $D=\{(m_0,m_1): m_0^2+m_1^2\le 1\}$ is the closed unit disk, whereas $S^1\!/\mathbb{Z}_2$, parametrized by the phase  $\eta\in[0,\pi[$, represents a circle with antipodal points identified, see the construction depicted in Fig.~\ref{fig:quotient}~(a).
The twist is due to the fact that, in order to identify the endpoints $\eta=0$ and $\eta=\pi$, the phase factor $\e^{\iu\pi}=-1\in\UU(1)$ has to be absorbed in the $\SU(2)$ part of $U=(M,\e^{\iu\eta})$ by sending $M\mapsto -M$, that is by sending $(m_0,m_1)\mapsto (-m_0,-m_1)$, see Fig.~\ref{fig:quotient}~(b). 

\begin{figure}[tb]
	\centering
	\includegraphics[trim=0 2.5cm 0 4cm, clip,
	width=\textwidth]{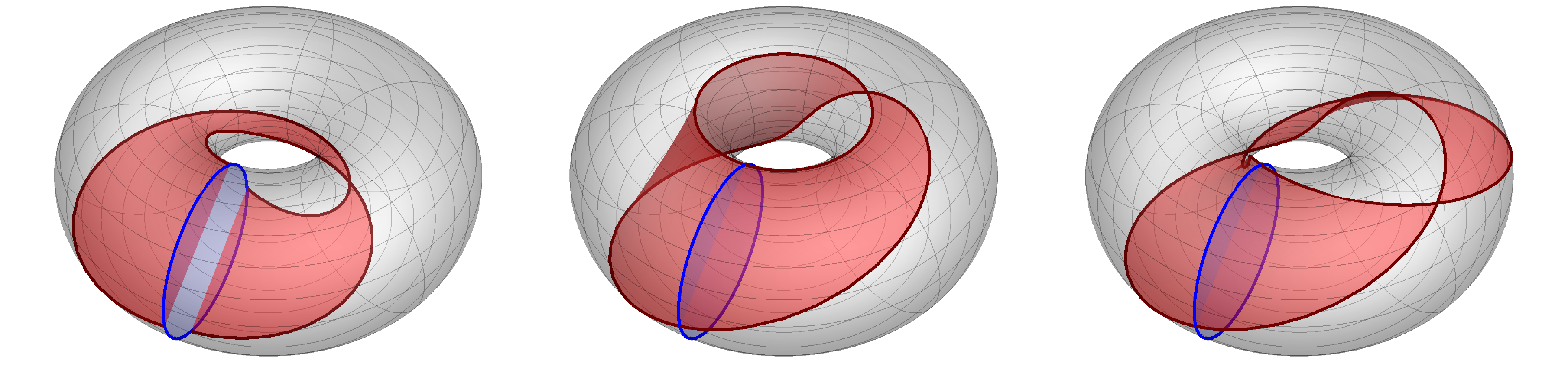}
	\caption{From left to right: solid tori respectively obtained by gluing the two bases of $D\times [0,\pi]$; by gluing the two bases of $D\times [0,\pi]$ after a global twist of $\pi$; and by gluing the two bases of $D\times [0,2\pi]$. In each plot the blue disk identifies the $\eta=0$ section, whereas the red surface represent the (families of) self-adjoint realizations which admit zero-modes, see Eq.~\eqref{eq:FU0}. The geometry of the spectral space $\Sigma$ is  the middle one, whose representation is both continuous and one-to-one.}
	\label{fig:tori}
\end{figure}

We remark that although a twisted torus is topologically indistinguishable from a non-twisted one, $S^1\!/\mathbb{Z}_2$ being indeed homeomorphic to $S^1$, a function defined on the torus is generally modified by the twist. The action of the twist, as well as its necessity, is highlighted in  Fig.~\ref{fig:tori}, where the twisted torus is compared with the non-twisted one and also with the ``doubled'' torus $\tilde{\Sigma}\cong D\times S^1$.

Let us now consider the space of Hamiltonians $\mathcal{M}$, in which two distinct self-adjoint realizations $H_U\neq H_{V}$, for $U\neq V$, always represent two distinct points. This space is given thus by the set
\begin{equation}
\mathcal{M}\equiv\{H_U \,:\, U\in\UU(2)\} \cong \UU(2)\,.
\end{equation} 
While each point of the twisted torus $\partial\Sigma\cong S^1\times (S^1\!/\mathbb{Z}_2)$ corresponds to exactly one of the $P$-symmetric Hamiltonians with quantum boundary conditions $U(\eta,\theta)$ of Eq.~\eqref{eq:Uparity}, each point of its interior $\mathring{\Sigma}=\Sigma\setminus\!\partial\Sigma$ is instead associated with an isospectral family of Hamiltonians characterized by the relation~\eqref{eq:PHP}, that is with an isospectral circle $S^1$ of self-adjoint realizations. We therefore obtain that
\begin{equation}
\mathcal{M}\cong \partial\Sigma\cup (\mathring{\Sigma}\times S^1)\,.
\end{equation}

\subsubsection{Role of time reversal.}
To conclude our analysis, we consider the role of the time reversal transformation 
\begin{equation}
  T\colon \psi(x)\mapsto\psi^{\ast}(x)
\end{equation}
 for $\psi\in L^2(-\ell /2,\ell /2)$, where $\psi^{\ast}(x)$ denotes the complex conjugate of $\psi(x)$. The action of the antiunitary operator $T$ on the Hamiltonian $H_U$ is given by
\begin{equation}
T H_U T^{\dagger}=H_{U^{\intercal}}\,,
\end{equation}
where $U^{\intercal}$ denotes the transpose of $U$. Since $U^{\intercal}$ can be obtained from $U$ simply by substituting $m_2$ with $-m_2$, we deduce that $T$ acts trivially on $\Sigma$: for each point $\sigma\in\Sigma$, represented e.g.\ by $\sigma=\sigma(H_U)$, we have that $\sigma(H_U)=\sigma(H_{U^{\intercal}})$. Moreover, since $U(\eta,\theta)=U^{\intercal}(\eta,\theta)$, the action of $T$ on   $\mathcal{M}$ is non-trivial only on the isospectral circle $S^1$, swapping each Hamiltonian having $m_2\neq 0$ with the corresponding Hamiltonian having $m'_2= -m_2$.

\section{Conclusions}
We can now come back to our initial question, whose answer is complete, but only partially positive: we can indeed hear the shape of \emph{some} boundary conditions. As we have shown, there is a family of boundary conditions (and thus of  Hamiltonians $H_U$) such that each element is uniquely associated with the corresponding spectrum $\sigma(H_U)$. These boundary conditions are described by the matrices $U(\eta,\theta)$ defined in Eq.~\eqref{eq:Uparity}, they belong to the \emph{boundary} of the spectral space $\Sigma$, and are associated with parity-symmetric self-adjoint realizations. Conversely, each spectrum $\sigma=\sigma(H_U)$ in the \emph{interior} of $\Sigma$ is associated with a $\UU(1)$-family of isospectral boundary conditions, given by Eq.~\eqref{eq:Udelta}. The corresponding Hamiltonians, which are not symmetric under parity, are related by Eq.~\eqref{eq:isospectrality} and thus all ``sound''  the same.

Moreover, we found that the spectral space $\Sigma$  has a non-trivial topology, being a twisted solid torus. This result, that more generally extends also to the space of Hamiltonians $\mathcal{M}$, is ultimately due to the non simply-connectedness of the unitary group $\UU(2)$  which parametrizes the quantum boundary conditions. In fact, an analogous ``M\"obius structure'' has also been found in a different setting~\cite{TFC01}. Future study can thus be devoted to explore the implications of this non-trivial topology, which we expect to be associated with spectral holonomies and geometric phases~\cite{ShWi89}. Also a concrete physical implementation of our model with  SQUIDs~\cite{Vion,Poletto,Paauw} or with Bose Einstein condensates in circular optical traps~\cite{Cataliotti} would  be of interest.

This work could further be expanded into different directions: by considering an analogous problem in higher-dimensional systems (which we expect to be a non-trivial task~\cite{extensions,trotter2}), as e.g.\ in a billiard~\cite{magnBill}, or by moving to more complex one-dimensional systems such as quantum graphs, or also by considering a relativistic setting described by the Dirac operator. In any case we believe that our results shed some light on the isospectrality problem of a ``quantum drum'' from a new point of view, which focuses on the boundary conditions rather than on the shape of the drum, and more generally contribute to the very active topic of boundary effects and their interplay with the bulk of a quantum system~\cite{AM12}.

\section*{Acknowledgments}
This work was partially supported by Istituto Nazionale di Fisic Nucleare (INFN) through the project ``QUANTUM'' and the Italian National Group of Mathematical Physics (GNFM-INdAM).

\appendix

\section{Group extensions}\label{sec:app}
In this Appendix, for the sake of completeness, we briefly review some basic facts about group extensions~\cite{Sco87}, and describe two relevant extensions involving the unitary group $\UU(2)$~\cite{AgSo00}.

For a triple of groups $N, E$ and $G$, connected by the homomorphisms $i\colon N\to E$ and $\pi\colon E\to G$, the sequence
\begin{equation}
1\to N\stackrel{i}{\to}E\stackrel{\pi}{\to}G\to 1\,,
\end{equation}
is said to be a short exact sequence as long as $i$ is injective, $\pi$ is surjective and $i(N)=\ker(\pi)$, where $\ker(\pi)=\{g\in E:\pi(g)=e_G\}$ and $e_G$ is the identity element of $G$. In this case, one can show that $i(N)$ is a normal subgroup of $E$ and that $E/i(N)\cong G\,,$ i.e.\ that $E$ is a \emph{group extension} of $G$ by $N$.

If moreover there exists a section, that is a homomorphism $s\colon G\to E$ satisfying $\pi\circ s=\id_{G}$, the extension $E$ is called \emph{split} and one can prove that it corresponds to the semidirect product $E=N\rtimes_\phi G$, where for each $g\in G$ the homomorphism $\phi_g\colon N\to N$ is given by
\begin{equation}
\phi_g(n)=i^{-1}\bigl(s(g)i(n)s(g^{-1})\bigr)\,.
\end{equation}
We recall that the composition law of $N\rtimes_\phi G$ is given by $(n,g) (n', g')=(n\phi_g(n'), gg')$. In particular, if $\phi_g=\id_N$ for each $g\in G$, the split extension is trivial, and $N\rtimes_\phi G$ is just the direct product $N\times G$.

\subsection{The unitary group $\UU(2)$}
There exist two interesting exact sequences involving $\UU(2)$. The first one is 
\begin{equation}
1\to \SU(2)\stackrel{i}{\hookrightarrow}\UU(2)\stackrel{\pi}{\to}\UU(1)\to 1\,,
\end{equation}
with $i$ the canonical injection and $\pi=\det$. Since the map
\begin{align}
s\colon \UU(1)\to \UU(2)\,,&&s(\e^{\iu\eta})=
\left(\begin{array}{@{}cc@{}} \e^{\iu\eta} &0 \\ 0 & 1
\end{array}\right)\,,
\end{align}
is a section (it is a homomorphism and $\pi\circ s=\id_{\UU(1)}$), we conclude that
\begin{align}
\UU(2)=\SU(2)\rtimes_\phi \UU(1)\,,&&\phi_{\e^{\iu\eta}}\colon 
\left(\begin{array}{@{}cc@{}} a & b \\ c & d \end{array}\right)
\mapsto
\left(\begin{array}{@{}cc@{}} a & \e^{\iu\eta}b \\ \e^{-\iu\eta}c & d \end{array}\right)\,.
\end{align}
%where 
%\begin{equation}
%\phi_{\e^{\iu\eta}}\colon 
%\left(\begin{array}{@{}cc@{}} a & b \\ c & d \end{array}\right)
%\mapsto
%\left(\begin{array}{@{}cc@{}} a & \e^{\iu\eta}b \\ \e^{-\iu\eta}c & d \end{array}\right)\,.
%\end{equation}
Using this construction,  a generic $\UU(2)$ element $U=(M,\e^{\iu\eta})$ can thus be represented by the matrix product of $M\in\SU(2)$ and of the section $s(\e^{\iu\eta})$:
\begin{equation}
U=Ms(\e^{\iu\eta})=M
\left(\begin{array}{@{}cc@{}} \e^{\iu\eta} &0 \\ 0 & 1
\end{array}\right)\,.
\end{equation}

The second exact short sequence is given instead by
\begin{equation}
1\to \Z_2\stackrel{i}{\to}\SU(2)\times \UU(1)\stackrel{\pi}{\to}\UU(2)\to 1\,,
\end{equation}
where now $i\colon \pm 1\mapsto (\pm \idm, \pm 1)$ and $\pi\colon (M, \e^{\iu\eta})\mapsto \e^{\iu\eta} M$. Also in this case we can find a section, namely
\begin{align}
s\colon \UU(2)\to \SU(2)\times \UU(1)\,,
&&s\colon U\mapsto \bigl(\det(U)^{-\frac{1}{2}}\,U, \det(U)^{\frac{1}{2}} \bigr)\,,
\end{align}
however the split extension turns out to be trivial, since $\phi_U=\id_{\Z_2}$ for each $U\in\UU(2)$ (another way to see this, after observing that $i(\Z_2)$ is a subgroup of the center $Z(\SU(2)\times\UU(1))=\Z_2\times \UU(1)$, is by recalling that each central extension which is split is also trivial). We can thus conclude that $\SU(2)\times\UU(1)$ is a \emph{double cover} of $\UU(2)$, in the sense that
\begin{equation}
\SU(2)\times\UU(1)=\Z_2\times \UU(2)\,,
\end{equation}
or equivalently that
\begin{equation}
\UU(2)=(\SU(2)\times\UU(1))/\Z_2\,.
\end{equation}
More concretely, by this second construction we can represent each $U\in\UU(2)$ by the projection $\pi((M,\e^{\iu\eta}))$, that is by
\begin{equation}
U=\e^{\iu\eta}M\,,
\end{equation}
\emph{as long as} we identify $(M,\e^{\iu\eta})$ with $(-M,-\e^{\iu\eta})=(-M,\e^{\iu(\eta+\pi)})$. As discussed in the main text, this identification gives rise to the twist of the spectral space.

\end{document}